# Many bioinformatics programming tasks can be automated with ChatGPT

Stephen R. Piccolo[1,*], Paul Denny[2], Andrew Luxton-Reilly[2], Samuel Payne[1], Perry G. Ridge[1]

1 - Department of Biology, Brigham Young University, Provo, UT, USA

2 - School of Computer Science, The University of Auckland, Auckland, New Zealand

* - Please address correspondence to S.R.P. at stephen_piccolo@byu.edu.

## Abstract

Computer programming is a fundamental tool for life scientists, allowing them to carry out many essential research tasks. However, despite a variety of educational efforts, learning to write code can be a challenging endeavor for both researchers and students in life science disciplines. Recent advances in artificial intelligence have made it possible to translate human-language prompts to functional code, raising questions about whether these technologies can aid (or replace) life scientists' efforts to write code. Using 184 programming exercises from an introductory-bioinformatics course, we evaluated the extent to which one such model—OpenAI's ChatGPT—can successfully complete basic- to moderate-level programming tasks. On its first attempt, ChatGPT solved 139 (75.5%) of the exercises. For the remaining exercises, we provided natural-language feedback to the model, prompting it to try different approaches. Within 7 or fewer attempts, ChatGPT solved 179 (97.3%) of the exercises. These findings have important implications for life-sciences research and education. For many programming tasks, researchers no longer need to write code from scratch. Instead, machine-learning models may produce usable solutions. Instructors may need to adapt their pedagogical approaches and assessment techniques to account for these new capabilities that are available to the general public.

## Introduction

For decades, the life-sciences community has called for researchers to gain a greater awareness of "computing in all its forms"[1]. This need is now greater than ever. A 2016 survey of principal investigators from diverse biology disciplines revealed that almost 90% of researchers had a current or impending need to use computational methods[2]. Computers can help researchers formalize scientific processes[3], accelerate research progress[4], improve job prospects, and even learn biological concepts[5]. These opportunities have motivated the creation of interdisciplinary training programs, courses, workshops, and tutorials to teach computing skills in a life-sciences context[6,2,@ 7,8–12]. In some circumstances, it is sufficient for researchers to understand computing concepts and learn to use existing tools; in other settings, learning to write computer code is invaluable[13]. A 2011 survey of scientists from many disciplines (other than computer science) found that researchers spent 35% of their time, on average, writing code[14]. Computer programming makes it possible to complete tasks not supported by existing tools, interface with software libraries, adapt algorithms based on custom needs, tidy data, and more[15–17]. In these applied scenarios, computer programs are often small[13], and the code might only be used for one particular project. Scripting languages are well suited to such tasks because researchers can focus on high-level needs and worry less about memory management, code efficiency, and other technical details[15]. Python, a scripting language, has gained much acceptance among scientists[14] and programming educators[18], perhaps due to its relatively simple syntax[19] and the availability of libraries supporting common tasks[20–24]. However, learning to program is a daunting challenge for many researchers. Decades of research have sought to characterize common errors and identify effective ways for novices to learn programming skills[25–30]; much remains to be discovered.

Recent advances in artificial intelligence have shown promise for converting natural-language descriptions of programming tasks to functional code[31,32]. The first such *large language models* (LLMs) fine-tuned to generate code that captured widespread interest were OpenAI's Codex and DeepMind's AlphaCode[33,34]. These models were trained on millions of code examples, representing diverse programming tasks. In November 2022, OpenAI released *ChatGPT*, an LLM that was fine-tuned with human feedback to generate natural dialogue-based text as well as code[35]. For basic to moderate-level programming tasks, these models may be able to aid researchers—or even replace their efforts. For more complicated projects, LLMs might be able to aid in writing or debugging portions of the code. If successful in these settings, LLMs could reduce the time that life scientists spend on programming, leaving more time for other research tasks.

We undertook a study to better understand the extent to which LLMs might be used in life-sciences research and education[36]. We used ChatGPT because it 1) was released recently, 2) is accessible with a Web browser, 3) has an ability to interact with users in a conversational manner, and 4) has garnered considerable attention among academics, industry competitors, and the general public[37–43]. In January 2023, ChatGPT had over 100 million active users[44].

We evaluated and documented ChatGPT's effectiveness on Python-programming exercises from an introductory-bioinformatics course taught primarily to undergraduates. We evaluated how well ChatGPT could interpret the prompts, respond to human feedback, and generate functional code. Here we describe quantitative and qualitative aspects of ChatGPT's performance, describe ways that ChatGPT could aid life scientists in research, and discuss implications for teaching and assessing students' programming capabilities in an educational context.

# Methods

*Programming exercises*

Since 2012, Brigham Young University has offered a course called *Introduction to Bioinformatics*. The course is intended to be accessible to students who are programming novices and have an interest in biology. One learning outcome for the course is that "students will be able to create computer scripts in the Python programming language to manipulate biological data stored in diverse file formats." To facilitate skill development, the instructors created Python programming exercises, which serve as formative assessments. When creating the exercises, we used online datasets, articles, and tools[45–54]. Six exercises in the second assignment were derived from an online course[55]; other than these, we are not aware that any of the exercises are in the public domain and thus could have been incorporated into ChatGPT's training data.

The exercises are organized into 19 assignments that each follow a particular theme. Students complete the assignments and exercises in a defined sequence. The assignments cover 1) relatively simple tasks like declaring and using variables, performing mathematical calculations, and writing conditional statements; 2) medium-difficulty tasks like working with strings, lists, loops, dictionaries, and files; and 3) more advanced tasks like writing regular expressions, manipulating tabular data, and creating data visualizations. Other assignments give students practice using techniques they learned in previous assignments. At specific times throughout the course, students complete additional programming exercises as summative assessments, culminating in an end-of-course assessment. We excluded the summative assessments from this study.

The programming exercises are delivered via a Web-based application that acts as an automatic grader. For each exercise, students receive a prompt describing the problem's context and requirements. In some cases, the prompt includes basic code that students can use as a starting point. For each exercise, at least one test is provided. For a given test, an input and an expected output are provided. The input may include data file(s); when applicable, these are provided within the prompt. The expected output may be text based ($n = 179$) or image based ($n = 5$). To generate the expected output, the instructor provides a solution, which accepts the input provided. When the same input is provided to a student's code, their output must match the expected output exactly. Students can make multiple attempts, as needed, without penalty. In many cases, instructors provide test(s) for which the inputs and/or expected outputs are hidden; this helps to prevent students from writing code that does not address the stated requirements. We excluded these tests from the study to maintain consistency with what students see; we manually verified whether ChatGPT-generated code met the requirements.

We used *openpyxl* (version 3.1.0)[56] to create a spreadsheet with information about each exercise. One column contains the prompt for each exercise, including the instructions and a summary of each test. Figure 1 shows an example of how the prompts were structured. For image-based tests, we did not include the expected outputs because ChatGPT does not accept images as input. In an attempt to make the prompts more understandable to ChatGPT—and to mimic what students or researchers might do—we added natural-language transitions between each section of the prompt. Other columns in the spreadsheet include the instructors' solutions and flags indicating whether each exercise was biology oriented. Many exercise prompts provide biology-based scenarios such that a basic understanding of biology concepts is helpful when interpreting the prompts.

*Evaluation approach*

We initiated a conversation with ChatGPT for each assignment. For one exercise at a time, we copied the prompt into ChatGPT's Web-based interface. To assess functional correctness, we copied ChatGPT's generated code into the automatic-grader that students use. If the code did not pass all of the tests, we continued the conversation with ChatGPT. In these interactions, we took the stance of a naive programmer who wishes to obtain functional code but is not necessarily able to provide feedback on the code itself. Accordingly, our feedback focused on helping ChatGPT understand the exercise requirements and was restricted to natural language such that no source code was present in any of our prompts. When we concluded that it might be helpful, we shortened the prompts. Other times we simply indicated that the previously generated solutions were unsuccessful and asked ChatGPT to try an alternative approach. When our interactions with ChatGPT suggested that a prompt lacked clarity, we slightly modified the prompt (and updated the spreadsheet accordingly). We allowed ChatGPT a maximum of ten attempts per exercise. As we interacted with ChatGPT, we used the spreadsheet to record the dates of the interactions, the ChatGPT version, ChatGPT's generated code (its final attempt), the number of passed tests, the number of attempts made by ChatGPT, and comments describing our interactions.

# Results

After filtering, 184 Python programming exercises were available for testing. ChatGPT successfully solved 139 (75.5%) of the exercises on its first attempt. When it did not, we engaged in a dialog with ChatGPT, focusing primarily on helping ChatGPT understand the requirements (see Methods). Of the remaining 45 exercises, ChatGPT solved 27 within 1 or 2 subsequent attempts. Within 7 or fewer attempts total, ChatGPT solved 179 (97.3%) of the exercises (Figure 2). As summarized in Table 1, the five unsolved exercises are delivered in the middle or end of the course; each requires students to combine multiple types of programming skills. One of these is the final exercise of the course; the instructors' solution uses 61 lines of code, nearly twice as many as any other solution. For the remaining exercises that failed, ChatGPT came close to passing the tests, but its solutions resulted in logic errors, runtime errors, or produced outputs that failed to match the expected outputs exactly.

We used statistics to understand more about the scenarios in which ChatGPT either succeeded or failed. First, we used the length of the instructors' solutions as an indicator for difficulty level. After removing comments (inline descriptions of how the code works) and blank lines, we compared the number of lines of code between the exercises that ChatGPT solved and those that it did not (Figure 3). The median for passing solutions was 6, and the median for non-passing solutions was 7; this difference was not statistically significant (Mann-Whitney U p-value: 0.2836). The lengths of the instructors' solutions showed a strong, positive correlation with the lengths of ChatGPT's solutions Figure 4, both for the number of characters (Spearman's rho = 0.89; p-value < 0.001) and the number of lines (Spearman's rho = 0.83; p-value < 0.001). Another indicator for difficulty level is the exercise-prompt length. For passing solutions, the median was 2019 characters, while the median was 9115 for non-passing solutions. Although this difference was not statistically significant (Mann-Whitney U p-value: 0.1021), it is consistent with a recent study of computer science exercises[32].

The number of attempts provides additional insight about ChatGPT's capabilities. Caution must be used when interpreting this metric because ChatGPT exhibits stochasticity. Whether it provides a correct answer on the first attempt or a later one, eventual success shows that its probabilistic model is capable of aiding users. However, a smaller number of attempts might suggest an ability to formulate a valid response more readily, thus requiring less time by the user. The number of attempts per exercise was significantly correlated with the length of the instructors' solution (rho = 0.234; p = 0.001) and the length of the prompt (rho = 0.31; p < 0.001). These correlations held, whether or not we considered the five exercises that ChatGPT failed to solve.

Of the 184 prompts, 98 (53.3%) were framed in a biological context. Of the five exercises that ChatGPT did not solve, four were framed in a biological context (Fisher's exact test p-value = 0.37). The biology-oriented prompts were longer than the remaining prompts (Mann-Whitney U p-value < 0.001). In the course, we frequently use biological data (e.g., genome sequences, medical observations, narrative text) to teach analysis skills and make the exercises more authentic. We included these data so that ChatGPT had an opportunity to recognize the files' structure. For 24 exercises, the prompt size exceeded the maximum allowed by ChatGPT. After we truncated the data to the first few lines, ChatGPT was successful at solving all of these exercises. On four other occasions, we shortened parts of the prompt as we interacted with ChatGPT in attempts to provide clarity. For example, we shortened the descriptions of how the code would be tested. ChatGPT was eventually successful on two of these four exercises.

We note additional challenges that ChatGPT faced when interpreting the programming prompts. On 21 exercises, ChatGPT used correct logic but produced outputs that were different from the expected outputs (for example, "Number of worms in the last box: 5" instead of "5"). Eventually, ChatGPT solved all of these exercises. On 25 exercises, ChatGPT generated code that produced logic errors; it eventually solved 20 of these exercises. On 10 exercises, ChatGPT generated code that produced runtime errors (exceptions); it eventually solved 8 of these exercises. On two exercises, ChatGPT generated passing code but the code did not directly address the prompt. For example, in one case, the prompt called for using a regular expression (text-based pattern matching), but ChatGPT used iteration logic instead; we marked these exercises as passing because the automatic grader did not verify which type of logic they used. On five occasions, as we interacted with ChatGPT, we noted parts of the prompt that may have been ambiguous. We clarified these prompts; subsequently, ChatGPT solved four of these exercises.

In using ChatGPT to solve these programming problems, we observed several practical issues that may impact the value offered by ChatGPT to researchers and students. For 13 exercises that did not pass on the first attempt, we asked ChatGPT to try an alternative approach and/or we re-delivered the original prompt. In these cases, we sought to take advantage of its stochastic nature, perhaps resulting in code that used a considerably different strategy. ChatGPT eventually passed 11 of these 13 exercises. For 41 (22.2%) exercises in total, ChatGPT used at least one programming technique that would *not* have been familiar to most students in the course. Many of these techniques are never taught in the course, whereas others are taught in later units. Finally, on six occasions, after an unsuccessful first attempt, ChatGPT generated code that did *not* address the original prompt. Conceivably, the model "forgot" earlier parts of the conversation or was "distracted" by subsequent inputs. Eventually, it solved all of these exercises.

## Discussion

These findings are remarkable and signal a new era for life scientists. For many basic- to moderate-level programming tasks, researchers no longer need to write code from scratch. Instead, they can rely on machine-learning models' abilities to translate natural-language descriptions to code. Researchers have already begun to explore this capability in practice[57]. We anticipate that as the models evolve, researchers and students will increasingly author programming prompts rather than code. Anecdotally, we found that this type of work is not always easy; during our evaluations, it was cognitively taxing at times to communicate with the model. In addition, these conversations were sometimes awkward. Although ChatGPT is capable of retaining a memory of previous interactions, its default response was to provide a solution; in many cases, it might have been more helpful for ChatGPT to request additional information or clarification regarding the problem. It was often more effective for us to restate the original prompt than engage in a back-and-forth dialog. On a positive note, ChatGPT was exceptionally effective at determining which parts of a given prompt were most informative; for example, it seemed to identify relevant aspects of biological context and ignore extraneous details. LLMs do not have an ability to execute code; thus, they are often not able to predict the output of code[58]. This is one area in which human feedback remains critical.

For 60+ years, researchers have been working to automate program synthesis[59,60]. Recent efforts have focused on training neural networks on large code repositories[33,34,61,62]. Our results show that ChatGPT represents a considerable advance compared to prior models. Chen, et al. evaluated Codex's ability to solve short- to medium-length programming exercises (median solution = 5.5 lines of code). When delivering the prompts, they used "docstrings" (structured descriptions of functions that must be generated). Codex was successful for 28.8% of these exercises in a single attempt; when making 100 attempts per exercise, it solved 77.5% of the exercises[33]. Austin, et al. used a different set of exercises that were either mathematical in nature or focused on core programming skills[58]. In contrast to Chen, et al., they used natural-language prompts (one or a few sentences). The solutions had a median length of 5 lines. Using various LLMs, they solved as many as 83.8% of the mathematical problems and 60% of the remaining problems (within 100 attempts). For a subset of the problems, they provided human-language feedback to the models (up to four interactions); maximum accuracy was 65%. In an educational context, Finnie-Ansley, et al. showed that Codex could solve 82.6% of programming exercises from an introductory, computer-science course within 10 attempts and that the model would have ranked among the top quartile of students in the course[31]. In a similar approach to ours, Denny et al. used Copilot (a development environment plug-in powered by OpenAI's Codex model) to solve 166 programming exercises designed for novice computer science students[63]. For problems that initially failed, they observed similar improvements in the performance of the model through natural-language modifications to the prompts. However, only 80% of the problems were ultimately solved. Given that the problems they analyzed were also designed for novices, the superior performance we observed may be due to improvements in the models themselves in the intervening six months.

Aside from the use of ChatGPT, our work differs from prior evaluations in scope and context. Previous studies used exercises that evaluated the models' abilities to solve mathematical problems or to use core programming skills like processing lists, processing strings, or evaluating integer sequences. Our exercises required similar techniques; but they also required the model to complete higher-level tasks like parsing data files, writing data files, creating graphics, and using external Python packages. Furthermore, more than half of our exercises were framed in a biological context. LLMs may be most useful for routine tasks that appear frequently in the training sets and only need to be modified for a particular purpose; however, our results show that LLMs can be used in new and diverse contexts as well.

Our findings have important implications for education. Until LLMs demonstrate an ability to replace all human programming efforts, it will remain necessary for students (and others) to gain programming skills[64]. In our course, it would be impossible to prevent students from using LLMs on formative assessments. However, summative assessments (exams) are the primary way that we determine students' final grades; these assessments are invigilated, and students do not have access to the Internet. Therefore, we retain confidence in the validity of grades determined under secure assessment conditions. Extensive practicing is a critical part of learning to write code[65,66]. Thus, if students rely on LLMs to generate answers to formative assessments without first devising their own solutions, they may be more likely to perform poorly on summative assessments. Indeed, over-reliance by novices was a key risk identified by Chen et al. when releasing the Codex model[33]. With the ease of use and wide availability of tools like Copilot and ChatGPT, novices may quickly learn to rely on auto-suggested solutions without thinking about the computational steps involved—or reading problem statements carefully. Furthermore, if students copy-and-paste code without understanding it—as has been observed for an online forum[67]— they may underperform on summative assessments. One way that instructors could counter this behavior is to use LLMs to generate student-specific questions about their code[68] to focus attention on code comprehension. As LLMs continue to mature and their use in code generation becomes more prevalent, it will become increasingly important for users of LLMs, both students and researchers, to be competent at code comprehension and code-evaluation skills. Educators will need to shift pedagogical practice to focus on understanding code that has been generated, and using rigorous testing to evaluate whether generated code meets specifications. LLMs provide opportunities to make learning processes more efficient. For example, when students attempt to devise solutions but become stuck, they may be able to use LLMs as intelligent tutors, for example to explain code fragments and related concepts[69–71]. Doing so may reduce the need for instructors or teaching assistants to provide help. Additionally, instructors can use LLMs when creating new exercises to evaluate whether their prompts are clear.

Our study has several limitations. We used one particular version of one LLM; thus, we do not know how our findings will generalize to other models, however it is likely that the performance of LLM-based code generators will continue to improve as model sizes increase. Our evaluation process was subjective by nature because it involved human judgment, and the human user was an expert instructor, not a student. Working with students to complete these evaluations might better reflect how ChatGPT is used in educational settings; however, students' strategies vary widely. One additional limitation is that the programming exercises we evaluated do not necessarily represent skills that would be taught in other introductory-bioinformatics courses or that would be used broadly in bioinformatics research. We used the Python programming language; our findings might not generalize to other languages. Future studies can shed additional light on how LLMs might be useful for bioinformatics research and education.

In this study, we provide evidence that dialogue-based LLMs, such as ChatGPT, can aid in solving programming exercises in the life sciences, particularly in the field of bioinformatics. However, despite generally excellent performance, these models cannot replace the need for human programming efforts entirely. In an authentic research setting, where an auto-grader is not available to provide instant feedback on the correctness of model-generated code, there is a risk that relying on their outputs may produce erroneous results. Nevertheless, the findings of this study have important implications for educators and researchers who seek to incorporate programming skills into their work. With the help of machine learning models, instructors can provide more personalized and efficient feedback to students, and researchers can accelerate their work by automating programming tasks.

# Table

**Table 1: Summary of exercises that ChatGPT did not solve.** ChatGPT failed to solve 5 of the exercises within 10 attempts. This table summarizes characteristics of these exercises and provides a brief summary of complications that ChatGPT faced when attempting to solve them.

| Assignment | Exercise | Prompt summary | Skills emphasized | Failure summary |
|---|---|---|---|---|
| 09 - Strings | 05 - Is CpG island | Count the proportion of a DNA sequence that consists of 'CG' nucleotide pairs | Manipulating strings; performing mathematical calculations | Failed on one test representing an edge case (exactly 10% of the sequence were CG pairs) |
| 14 - Regular Expressions 1 | 06 - Find words that start with a vowel | Find words in a biological text that begin with vowels | Writing regular expressions; reading files | Trouble dealing with extra spaces or punctuation marks |
| 15 - Regular Expressions 2 | 06 - Switch column order | Switch the first two columns in a tab-delieted text file | Writing regular expressions or using lists; reading files; writing files | Runtime errors, logic errors dealing with + or - portion of blood types |
| 19 - Additional Practice | 08 - Make inducible promoter - Part B | Identify in-frame start and stop codons in an mRNA sequence | Manipulating strings; reading files; using complex iteration logic | Failed to follow the instruction to look for the start codon in frame |
| 19 - Additional Practice | 10 - Make inducible promoter - Part D | Identify restriction-enzyme binding sites upstream of a gene in a DNA sequence | Reading files; writing regular expressions; using complex iteration logic; using lists; using dictionaries; using conditionals | Runtime errors, various logic errors, failure to fully comprehend the prompt |

# Figures

```
Here is a new Python programming task.

Please write a Python function called "addNumbers" that calculates the sum of two numeric values. The functio
should accept two arguments (the two numbers to be summed) and return the sum of these values.

Below is a description of a test to verify your code.

For this test, the following code will be executed after your code:

    print(addNumbers(2, 2))

Below is the expected output:

    4

Below is a description of a test to verify your code.

For this test, the following code will be executed after your code:

    print(addNumbers(0, 0))

Below is the expected output:

    0

Below is a description of a test to verify your code.

For this test, the following code will be executed after your code:

    print(addNumbers(-1, 5001))

Below is the expected output:

    5000

Below is a description of a test to verify your code.

For this test, the following code will be executed after your code:

    print(addNumbers(-5000.0, 5000.0))

Below is the expected output:

    0.0

Below is a description of a test to verify your code.

For this test, the following code will be executed after your code:

    print(addNumbers(5.5, 1))

Below is the expected output:

    6.5
```

**Figure 1: Example prompt for a programming exercise delivered to ChatGPT.**

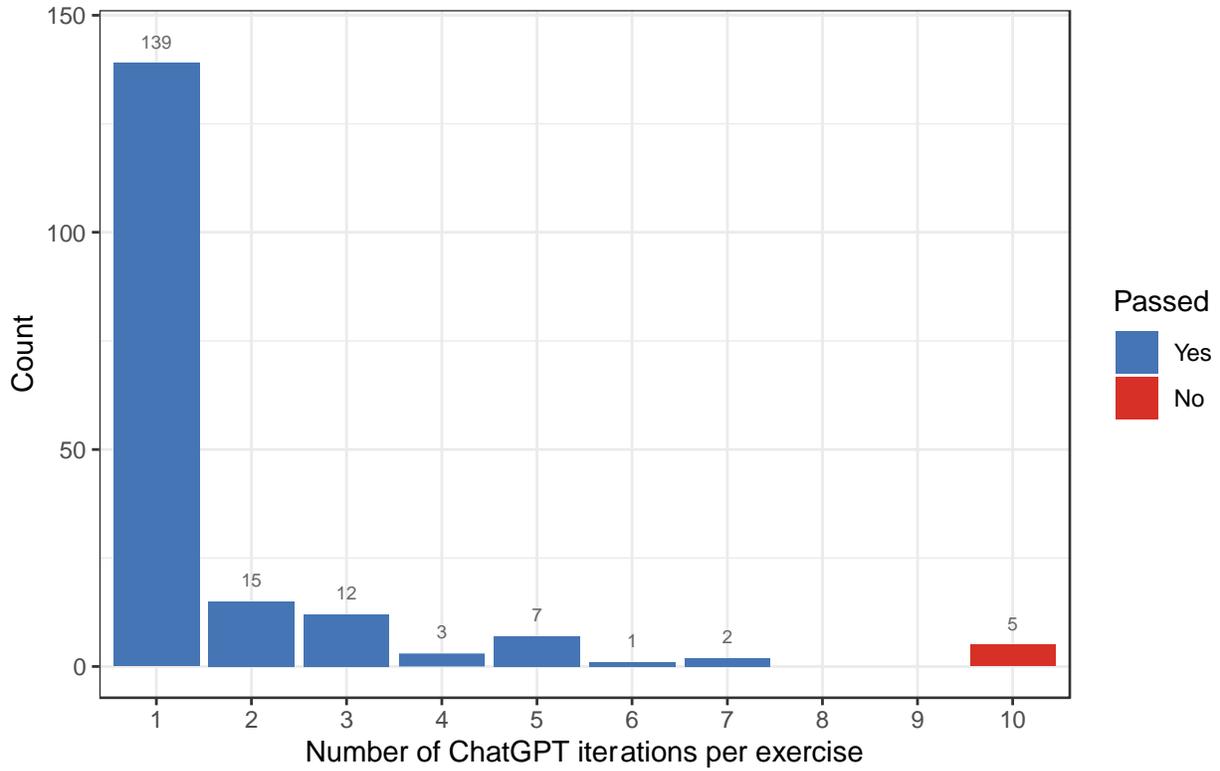

**Figure 2: Number of ChatGPT iterations per exercise.** For each exercise prompt, we gave ChatGPT up to 10 attempts at generating a code solution that successfully passed the tests. The counts above each bar represents the number of exercises that required a particular number of attempts.

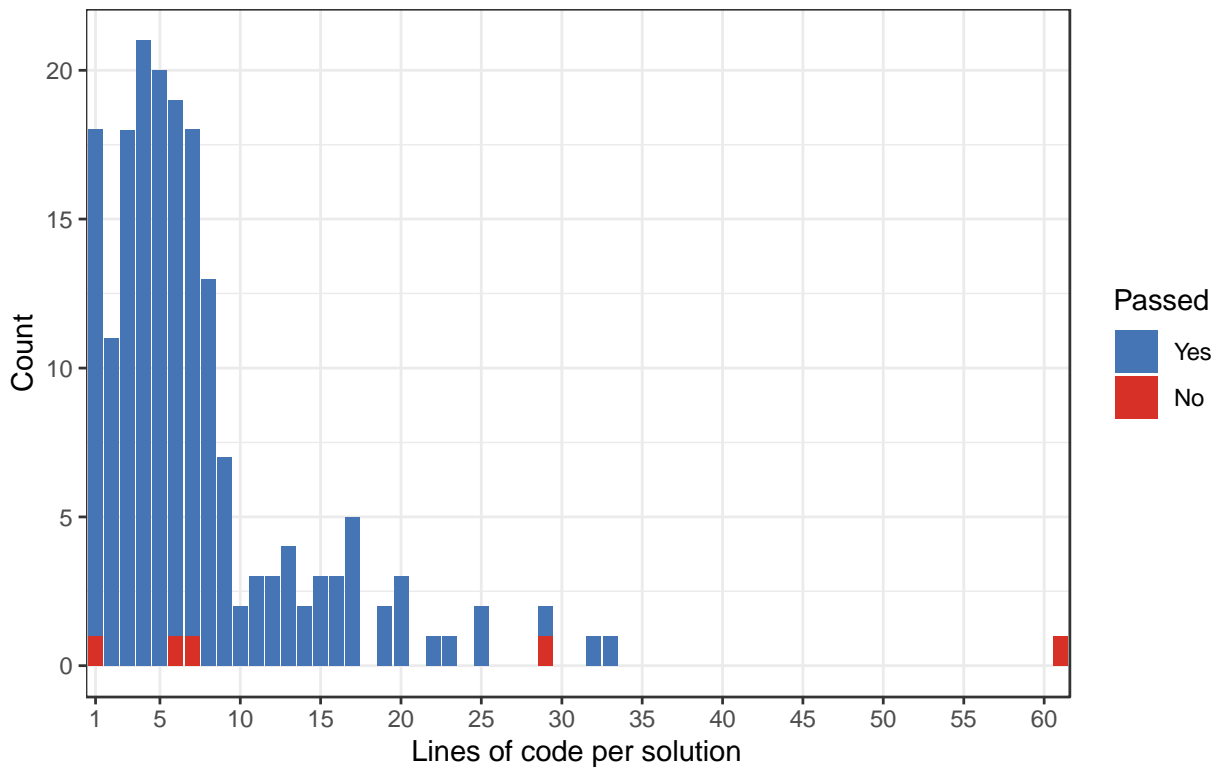

**Figure 3: Lines of Python code per instructor solution.** Course instructors provided a solution for each exercise. This plot illustrates the number of lines of code for each solution, after removing comment lines.

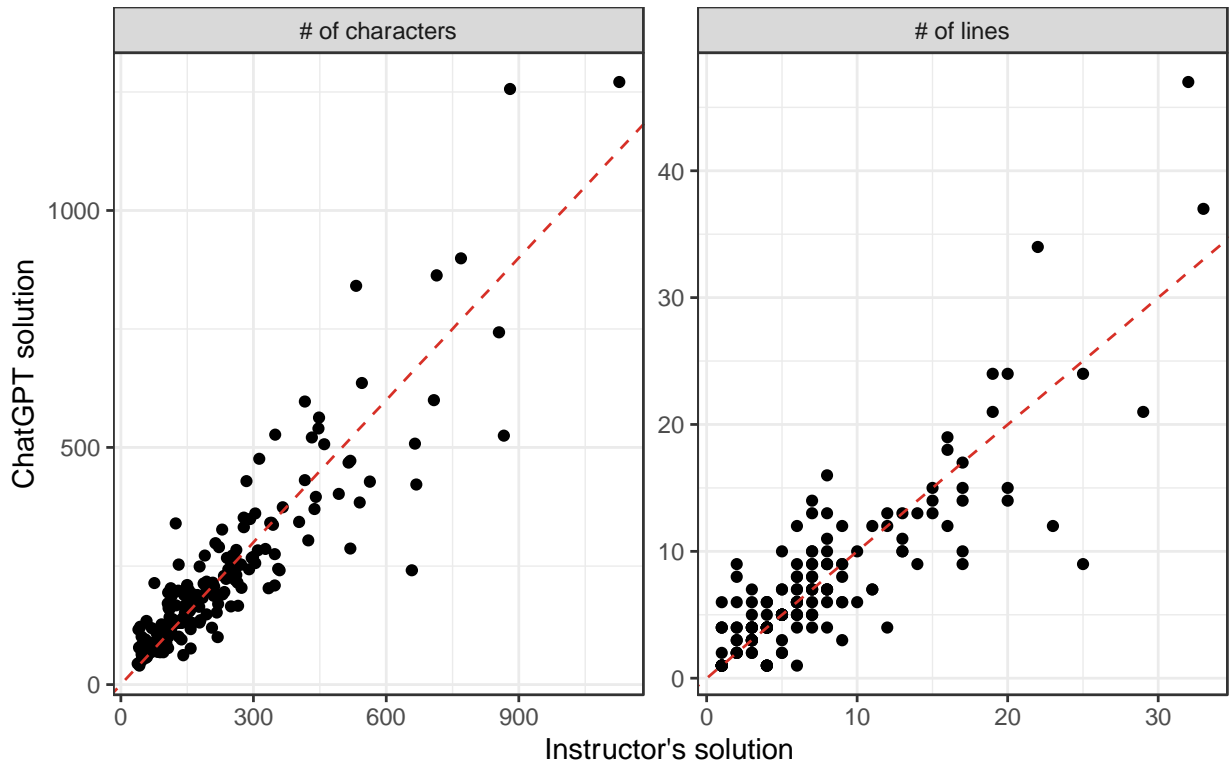

**Figure 4: Comparison of code-solution lengths for instructor solutions versus ChatGPT solutions.** This illustrates the relationship between A) the number of characters or B) the number of lines of code, for each exercise, after removing comment lines. The dashed, red line is the identity line.